\documentclass[10pt,twoside,twocolumn]{IEEEtran}%
\usepackage{amssymb}
\usepackage{epsfig,graphicx,subfigure,amsmath,nicefrac}
\usepackage{times}%
\usepackage{amsmath}%
\usepackage{amsfonts}%
\usepackage{graphicx}
\providecommand{\U}[1]{\protect\rule{.1in}{.1in}}
\newtheorem{theorem}{Theorem}[section]
\newtheorem{corollary}[theorem]{Corollary}
\newtheorem{lemma}[theorem]{Lemma}
\newtheorem{proposition}[theorem]{Proposition}

\newtheorem{definition}[theorem]{Definition}
\newtheorem{remark}[theorem]{Remark}

\newcommand\nc\newcommand
\nc\bfa{{\mathbf a}}
\nc\bfA{{\mathbf A}}\nc\cA{{\mathcal A}}
\nc\bfb{{\mathbf b}}\nc\bfB{{\mathbf B}}\nc\cB{{\mathcal B}}
\nc\bfc{{\mathbf c}}\nc\bfC{{\mathbf C}}\nc\cC{{\mathcal C}}
\nc\bfd{{\mathbf d}}\nc\bfD{{\mathbf D}}\nc\cD{{\mathcal D}}
\nc\bfe{{\mathbf e}}\nc\bfE{{\mathbf E}}\nc\cE{{\mathcal E}}
\nc\bff{{\mathbf f}}\nc\bfF{{\mathbf F}}\nc\cF{{\mathcal F}}
\nc\bfg{{\mathbf g}}\nc\bfG{{\mathbf G}}\nc\cG{{\mathcal G}}
\nc\bfh{{\mathbf h}}\nc\bfH{{\mathbf H}}\nc\cH{{\mathcal H}}
\nc\bfi{{\mathbf i}}\nc\bfI{{\mathbf I}}\nc\cI{{\mathcal I}}
\nc\bfj{{\mathbf j}}\nc\bfJ{{\mathbf J}}\nc\cJ{{\mathcal J}}
\nc\bfk{{\mathbf k}}\nc\bfK{{\mathbf K}}\nc\cK{{\mathcal K}}
\nc\bfl{{\mathbf l}}\nc\bfL{{\mathbf L}}\nc\cL{{\mathcal L}}
\nc\bfm{{\mathbf m}}\nc\bfM{{\mathbf M}}\nc\cM{{\mathcal M}}
\nc\bfn{{\mathbf n}}\nc\bfN{{\mathbf N}}\nc\cN{{\mathcal N}}
\nc\bfo{{\mathbf o}}\nc\bfO{{\mathbf O}}\nc\cO{{\mathcal O}}
\nc\bfp{{\mathbf p}}\nc\bfP{{\mathbf P}}\nc\cP{{\mathcal P}}
\nc\bfq{{\mathbf q}}\nc\bfQ{{\mathbf Q}}\nc\cQ{{\mathcal Q}}
\nc\bfr{{\mathbf r}}\nc\bfR{{\mathbf R}}\nc\cR{{\mathcal R}}
\nc\bfs{{\mathbf s}}\nc\bfS{{\mathbf S}}\nc\cS{{\mathcal S}}
\nc\bft{{\mathbf t}}\nc\bfT{{\mathbf T}}\nc\cT{{\mathcal T}}
\nc\bfu{{\mathbf u}}\nc\bfU{{\mathbf U}}\nc\cU{{\mathcal U}}
\nc\bfv{{\mathbf v}}\nc\bfV{{\mathbf V}}\nc\cV{{\mathcal V}}
\nc\bfw{{\mathbf w}}\nc\bfW{{\mathbf W}}\nc\cW{{\mathcal W}}
\nc\bfx{{\mathbf x}}\nc\bfX{{\mathbf X}}\nc\cX{{\mathcal X}}
\nc\bfy{{\mathbf y}}\nc\bfY{{\mathbf Y}}\nc\cY{{\mathcal Y}}
\nc\bfz{{\mathbf z}}\nc\bfZ{{\mathbf Z}}\nc\cZ{{\mathcal Z}}
\newcommand{\remove}[1]{}
\nc\bsa{{\boldsymbol a}}\nc\bsA{{\boldsymbol A}}
\nc\bsb{{\boldsymbol b}}\nc\bsB{{\boldsymbol B}}
\nc\bsc{{\boldsymbol c}}\nc\bsC{{\boldsymbol C}}
\nc\bsd{{\boldsymbol d}}\nc\bsD{{\boldsymbol D}}
\nc\bse{{\boldsymbol e}}\nc\bsE{{\boldsymbol E}}
\nc\bsf{{\boldsymbol f}}\nc\bsF{{\boldsymbol F}}
\nc\bsg{{\boldsymbol g}}\nc\bsG{{\boldsymbol G}}
\nc\bsh{{\boldsymbol h}}\nc\bsH{{\boldsymbol H}}
\nc\bsi{{\boldsymbol i}}\nc\bsI{{\boldsymbol I}}
\nc\bsj{{\boldsymbol j}}\nc\bsJ{{\boldsymbol J}}
\nc\bsk{{\boldsymbol k}}\nc\bsK{{\boldsymbol K}}
\nc\bsl{{\boldsymbol l}}\nc\bsL{{\boldsymbol L}}
\nc\bsm{{\boldsymbol m}}\nc\bsM{{\boldsymbol M}}
\nc\bsn{{\boldsymbol n}}\nc\bsN{{\boldsymbol N}}
\nc\bso{{\boldsymbol o}}\nc\bsO{{\boldsymbol O}}
\nc\bsp{{\boldsymbol p}}\nc\bsP{{\boldsymbol P}}
\nc\bsq{{\boldsymbol q}}\nc\bsQ{{\boldsymbol Q}}
\nc\bsr{{\boldsymbol r}}\nc\bsR{{\boldsymbol R}}
\nc\bss{{\boldsymbol s}}\nc\bsS{{\boldsymbol S}}
\nc\bst{{\boldsymbol t}}\nc\bsT{{\boldsymbol T}}
\nc\bsu{{\boldsymbol u}}\nc\bsU{{\boldsymbol U}}
\nc\bsv{{\boldsymbol v}}\nc\bsV{{\boldsymbol V}}
\nc\bsw{{\boldsymbol w}}\nc\bsW{{\boldsymbol W}}
\nc\bsx{{\boldsymbol x}}\nc\bsX{{\boldsymbol X}}
\nc\bsy{{\boldsymbol y}}\nc\bsY{{\boldsymbol Y}}
\nc\bsz{{\boldsymbol z}}\nc\bsZ{{\boldsymbol Z}}

\nc{\veps}{\varepsilon} \nc{\ds}{\displaystyle}
\nc{\ts}{\textstyle}
\begin{document}

\title{On the Fingerprinting Capacity Under the Marking Assumption}
\author{N.~Prasanth~Anthapadmanabhan, Alexander~Barg,~\IEEEmembership{Fellow,~IEEE,} 
and~Ilya~Dumer,~\IEEEmembership{Fellow,~IEEE}\thanks{%
N.~P.~Anthapadmanabhan is with the Department of Electrical and Computer
Engineering, University of Maryland, College Park, MD 20742 USA (e-mail:
nagarajp@umd.edu). Supported in part by NSF grant CCF0515124.} \thanks{
A.~Barg is with the Department of Electrical and Computer Engineering and
Institute for Systems Research, University of Maryland, College Park, MD
20742 USA (e-mail: abarg@umd.edu). Supported in part by NSF grants
CCF0515124, CCF0635271 and by NSA grant H98230-06-1-0044.} \thanks{%
I. Dumer is with the Department of Electrical Engineering, University of
California, Riverside, CA 92521 USA (e-mail: dumer@ee.ucr.edu). Supported in
part by NSF grants CCF0622242 and CCF063533.}}
\maketitle

\begin{abstract}
We address the maximum attainable rate of fingerprinting codes
under the marking assumption, studying lower and upper bounds on
the value of the rate for various sizes of the attacker coalition.
Lower bounds are obtained by considering typical coalitions, which
represents a new idea in the area of fingerprinting and enables us
to improve the previously known lower bounds for coalitions of
size two and three. For upper bounds, the fingerprinting problem
is modelled as a communications problem. It is shown that the
maximum code rate is bounded above by the capacity of a certain
class of channels, which are similar to the multiple-access
channel. Converse coding theorems proved in the paper provide new
upper bounds on fingerprinting capacity.

It is proved that capacity for fingerprinting against coalitions of size two
and three over the binary alphabet satisfies $0.25\leq C_{2,2}\leq0.322$ and
$0.083\leq C_{3,2}\leq0.199$ respectively. For coalitions of an arbitrary
fixed size $t,$ we derive an upper bound $(t\ln2)^{-1}$ on fingerprinting
capacity in the binary case$.$ Finally, for general alphabets, we establish
upper bounds on the fingerprinting capacity involving only single-letter
mutual information quantities.

\end{abstract}

\markboth{IEEE TRANSACTIONS ON INFORMATION THEORY}
{Anthapadmanabhan \MakeLowercase{\textit{et al.}}:
Fingerprinting Capacity Under the Marking Assumption}


%\IEEEpeerreviewmaketitle

\section{Introduction}

\label{sect:cap-intro}

\PARstart{T}{he} distribution of licensed digital content (e.g., software,
movies, music etc.) has become increasingly popular in recent times. With this
comes the need to protect the copyright of the distributor against
unauthorized redistribution of the content (\emph{piracy}).

To introduce the problem, we begin with an informal description. Suppose the
distributor has some content which he would like to distribute among a set of
licensed users. One can think of a simple scheme where each licensed copy is
identified by a unique mark (\emph{fingerprint}) which is embedded in the
content and is imperceptible to the users of the system. Note that the
distributed copies are identical except for the fingerprints. If a naive user
distributes a copy of his fingerprinted content, then the pirated copy can
easily be traced back to the guilty user and hence he will be exposed. Tracing
the guilty user becomes more difficult when a collection of users
(\emph{pirates}) form a \emph{coalition} to detect the fingerprints and
modify/erase them before illegally distributing the data. Digital
fingerprinting is a technique that assigns to each user a mark in a way that
enables the distributor to identify at least one of the members of the
coalition as long as its size does not exceed a certain threshold $t$, which
is a parameter of the problem.

There are two main setups considered for the fingerprinting problem in the
literature. The \emph{distortion} setting is commonly used in applications
relating to multimedia fingerprinting \cite{moul-sull, som-mer}. In this
model, the fingerprint is usually a ``covert signal'' which is superimposed on
the original ``host'' data in such a way that the difference, or distortion,
between the original and the fingerprinted copies is smaller than some
threshold. The coalitions are restricted to creating a forgery which has
distortion less than some threshold from at least one of the colluders'
fingerprinted copies.

On the other hand, we have the \emph{marking assumption} setting introduced in
\cite{boneh} which will be our main interest in this paper. In this case, the
fingerprint is a set of redundant digits which are distributed in some random
positions (unknown to the users) across the information digits of the original
content. The fingerprint positions remain the same for all users. It is
assumed that these redundant digits do not affect the functionality of the
content, while tampering with an information digit damages the content
permanently. The motivation for this assumption comes from applications to
software fingerprinting, where modifying arbitrary digits can damage its functionality.

The coalition attempts to discover some of the fingerprint positions by
comparing their marked copies for differences. If they find a difference in
some position, it is guaranteed to be a redundant fingerprint digit. In the
other positions, it could be either an information digit or a fingerprint
digit. The \emph{marking assumption} states that the coalitions may modify
only those positions where they find a difference in their fingerprinted
copies. Hence, in analyzing this model, it becomes sufficient to just look at
the fingerprint positions and ignore the information digits. The collection of
fingerprints distributed to all the users of the system together with the
strategy of decoding (pirate identification) used is called a \emph{code}
below. A code is said to be \emph{$t$-fingerprinting} or
\emph{collusion-secure} against coalitions of $t$ pirates if the error
probability of decoding approaches 0 as the code length tends to $\infty.$

Collusion-secure fingerprinting codes were introduced by Boneh and Shaw
\cite{boneh}. It was shown in \cite{boneh} that for any single deterministic
code, the probability of decoding error in the ``wide-sense'' formulation (see
Section \ref{sect:cap-def}) is bounded away from zero. Hence, it becomes
necessary for the distributor to use some form of randomization, where the
random key is known only to the distributor, in order to construct such
fingerprinting codes. This paper also gave the first example of codes with
vanishing error probability. Further general constructions were proposed by
Barg et al. \cite{bbk}
%which presented a family of codes of exponential size
%and error  probability falling exponentially with the code
%length.
and Tardos \cite{tardos}.

The case of zero error probability was considered independently by Hollmann et
al. \cite{hol98} who termed them as codes with the \emph{identifiable parent
property}, or IPP codes. They were further studied in \cite{hypipp, alon-stav,
blackburn-ub,stad-stin-wei} among others.

In this paper, we are interested in computing the fundamental limits of the
fingerprinting problem, i.e., in establishing bounds on the capacity (or
maximum attainable rate) of fingerprinting codes. We denote by $C_{t,q}$ the
capacity of fingerprinting with $q$-ary codes against coalitions of size $t$
(this quantity is defined formally later in the paper). The problem of
determining the fingerprinting capacity was raised in \cite{bbk}. To date,
only some lower bounds are known through constructions and existence results:
$C_{2,2}\ge0.2075$ \cite{kaba}; $C_{3,2}\ge0.064$ \cite{p-barg}; $C_{t,2}%
\ge(100 t^{2} \ln2 )^{-1}, t\ge2$ \cite{tardos}.

New capacity bounds of our paper are based on an information-theoretic view of
the fingerprinting problem. They are established as follows. Attainability
results (lower bounds) are shown by random coding techniques which take into
account the typical coalitions, i.e., the coalitions that occur with high
probability. This represents a new idea in fingerprinting which enables us to
improve random choice arguments of various kinds used earlier in
\cite{p-barg,boneh,kaba,tardos}. For upper bounds we model fingerprinting as a
multi-user communications channel. A converse theorem for a transmission
scenario that models some aspects of the fingerprinting problem is proved to
establish an upper bound on the capacity of fingerprinting.

It should be noted that a similar information-theoretic approach to finding
the capacity of fingerprinting was previously studied in \cite{som-mer} and
\cite{ahls-ipp}. In \cite{som-mer}, the authors obtain upper and lower bounds
on the capacity of
fingerprinting with distortion constraints as opposed to the marking
assumption setting of this paper.
Paper \cite{ahls-mac} uses the marking assumption setting, but it
addresses a simpler problem whose results do not directly apply to 
fingerprinting.

The rest of the paper is organized as follows. In Section \ref{sect:cap-def},
we recall the statement of the fingerprinting problem and give an
information-theoretic formulation. We also prove several results related to
the problem statement that justify various techniques used to derive bounds on
the fingerprinting capacity later in the paper. In particular, lower bounds on
$C_{t,2}, t=2,3$ are proved in Section \ref{sect:cap-lb2}. Sections
\ref{sect:cap-ub-w} and \ref{sect:cap-ub-s} are devoted to upper bounds on
$C_{t,q}$ for arbitrary $t,q$ and their specializations for $t=2,3$ in the
case of the binary alphabet.

\section{Problem statement}

\label{sect:cap-def}

\subsection{Notation}

Random variables (r.v.'s) will be denoted by capital letters and their
realizations by low-case letters. The probability distribution of a r.v. $X$
will be denoted by $P_{X}.$ If $X$ and $Y$ are independent r.v.'s, then their
joint distribution is written as $P_{X} \times P_{Y}.$ For positive integers
$l,m,$ $X_{l}^{l+m}$ will denote the collection of r.v.'s $\{X_{l}%
,X_{l+1},\dots,X_{l+m}\},$ and the shorthand $[l]$ will be used to denote the
set $\{1,\dots,l\}.$ Boldface will denote vectors of length $n.$ For example,
$\bsx$ denotes a vector $(x_{1},\dots,x_{n})$ and $\bsX$ denotes a random
vector $(X_{1},\dots,X_{n}).$ The Hamming distance between vectors $\bsx,\bsy$
will be written as $\qopname\relax{no}{dist}(\bsx,\bsy).$ We will denote the
binary entropy function by $h(x):= -x \log_{2} x -(1-x)\log_{2}(1-x)$ and
$1(\cdot)$ will represent the indicator function.

\subsection{Fingerprinting codes}

Let $\cQ$ denote an alphabet of (finite) size $q.$ Let $M$ be the number of
users in the system and let $n$ denote the length of the fingerprints. Assume
that there is some ordering of the users and denote their set by
$\cM=\{1,\dots,M\}$. Let $\cK$ be a finite set whose size may depend on $n.$
Elements of the set $\cK$ will be called keys. For every $k \in\cK,$ let
$(f_{k},\phi_{k})$ be an $n$-length \emph{code}, i.e., a pair of encoding and
decoding mappings:
\begin{equation}
\label{eqn:intro-f}f_{k}: \cM \to\cQ^{n}%
\end{equation}
\begin{equation}
\label{eqn:intro-phi}\phi_{k}: \cQ^{n} \to\cM \cup\{0\}
\end{equation}
where the decoder output 0 will signify a decoding failure. By definition, the
fingerprinting system is formed by a \emph{randomized code}, i.e., a random
variable $(F,\Phi)$ taking values in the family $\{(f_{k},\phi_{k}), k
\in\cK \}$. Note that the dependence on $n$ has been suppressed in this
notation for simplicity. The \emph{rate} of this code is $R=n^{-1}\log_{q} M$.

The system operates as follows. The distributor chooses a key $k$ according to
a probability distribution $\pi(k)$ on $\cK$ and assigns the fingerprint
$f_{k}(i)$ to user $i$. On receiving a forged fingerprint, the distributor
uses the tracing strategy $\phi_{k}$ (corresponding to the selected key) to
determine one of the guilty users.

We need randomization because: (a) deterministic fingerprinting codes do not
exist in certain formulations \cite{boneh,bbk}, and (b) we allow the family of
encoders and decoders and the distribution $\pi(k)$ to be known to all users
of the system. The only advantage the distributor has is the knowledge of the
particular key being used. This assumption follows the accepted standards of
cryptographic systems where it is usually assumed that the
encryption/decryption algorithms are publicly available and that the only
parameter kept secret by the system's constructor is the key.

The fingerprints are assumed to be distributed inside the host message so that
its location is unknown to the users. The location of the fingerprints,
however, remains the same for all users. A \emph{coalition} $U$ of $t$ users
is an arbitrary $t$-subset of $\{1,\dots,M\}.$ Following accepted usage, we
will refer to the members of the coalition as ``pirates''. The coalition
observes the collection of their fingerprints $f_{k}(U)=\{\bsx_{1}%
,\dots,\bsx_{t}\}$ and attempts to create a fingerprint $\bsy \in\cQ^{n}$ that
does not enable the distributor to trace it back to any of the users in $U$.
Note that although the fingerprint locations are not available to the pirates,
they may attempt to detect some of these locations by comparing their copies
for differences. Thus, coordinate $i$ of the fingerprints is called
\emph{undetectable} for the coalition $U$ if
\[
x_{1i}=x_{2i}=\dots=x_{ti}%
\]
and is called \emph{detectable} otherwise.

\medskip

\begin{definition}
The \emph{marking assumption} states that for any fingerprint $\bsy$ created
by the coalition $U$, $y_{i}=x_{1i}=x_{2i}=\dots=x_{ti}$ in every coordinate
$i $ that is undetectable.
\end{definition}

\medskip

In other words, in creating $\bsy$, the pirates can modify only detectable positions.

For a given set of observed fingerprints $\{\bsx_{1},\dots,\bsx_{t}\},$ the
set of forgeries that can be created by the coalition is called the
\emph{envelope}. Its definition depends on the exact rule the coalition should
follow to modify the detectable positions:

\begin{itemize}
\item If the coalition is restricted to use only a symbol from their assigned
fingerprints in the detectable positions, we obtain the \emph{narrow-sense
envelope}:
\begin{equation}
\label{eqn:intro-nenv}\cE_{N}(\bsx_{1},\dots,\bsx_{t})= \{\bsy \in\cQ^{n} |
y_{i} \in\{x_{1i},\dots,x_{ti}\}, \forall i \};
\end{equation}

\item If the coalition can use any symbol from the alphabet in the detectable
positions, we obtain the \emph{wide-sense envelope}:
\begin{equation}
\label{eqn:intro-wenv}\cE_{W}(\bsx_{1},\dots,\bsx_{t})= \{\bsy \in\cQ^{n} |
y_{i}=x_{1i}, \forall i \text{ undetectable}\}.
\end{equation}

\end{itemize}

We remark that there are further generalizations of the rules above where
coalitions are also allowed to erase the symbols in detectable positions. This
generalization is not considered below; we refer the interested reader to
\cite{bbk}. In the following, we will use $\cE(\cdot)$ to denote the envelope
from any of the rules or their generalizations mentioned above.

\begin{remark}
\label{rem:intro-bin} The definition for a fingerprinting code depends on the
envelope considered. Therefore,  different problems can arise for each
definition of the envelope. The binary alphabet is of special interest because
of its wide use in practical digital applications. For this special case, it
is easy to see that the narrow-sense and wide-sense envelopes are exactly the same.
\end{remark}

Suppose that the coalition $U$ uses a randomized strategy $V(\cdot|\cdot
,\dots,\cdot)$ to create the new fingerprint, where $V(\bsy|\bsx_{1}%
,\dots,\bsx_{t})$ gives the probability that the coalition creates $\bsy$
given that it observes the fingerprints $\bsx_{1},\dots,\bsx_{t}$. A strategy
$V$ is called admissible if
\[
V(\bsy|\bsx_{1},\dots,\bsx_{t})>0 \text{ only if } \bsy \in\cE(\bsx_{1}%
,\dots,\bsx_{t}).
\]
Let $\cV_{t}$ denote the class of admissible strategies. Such randomized
strategies model any general attack the coalition is capable of and also
facilitate mathematical analysis. The distributor, on observing the suspect
fingerprint $\bsy$, uses the decoder $\phi_{k}$ while using the key $k$. Then
the probability of error for a given coalition $U$ and strategy $V$ averaged
over the family of codes is defined as follows:
\begin{equation}
\label{eqn:intro-pestr}e(U, F, \Phi, V)= \mathsf{E}_{K} \sum_{\substack{\bsy:
\\\phi_{K}(\bsy) \notin U}} V(\bsy|f_{K}(U))
\end{equation}
where $\mathsf{E}_{K}$ is the expectation with respect to the distribution
$\pi(k).$

\begin{definition}
\label{def:cap-fingcode} A randomized code $(F,\Phi)$ is said to be
$t$-\emph{fingerprinting with $\varepsilon$-error} if
\begin{equation}
\label{eqn:cap-fingdef}\max_{V \in\cV_{\tau}} \max_{U:|U|= \tau} e(U,F,\Phi,V)
\le\veps, \quad\forall\tau\le t.
\end{equation}

\end{definition}

\subsection{Fingerprinting capacity}

\label{sect:cap-itform}

We now formulate the fingerprinting problem as a communications problem in
which the set of messages is identified with the set of users of the
fingerprinting system. Each message is mapped to a codeword which corresponds
to the fingerprint of the user. Any set of $t$ messages (a coalition) may be
chosen, and they are transmitted over an \emph{unknown} $t$%
-input-single-output channel defined by the strategy of the coalition. The
class of possible channels will be defined by the marking assumption. The
output of the channel (that represents the strategy) gives the forged
fingerprint. The task of the decoder is to recover at least \emph{one} of the
transmitted messages to have produced the channel output.

Observe that this information-theoretic model differs from the traditional
$t$-user Multiple-Access Channel (MAC) because: (a) the decoder makes an error
only when its output does not match \emph{any of the transmitted messages},
and (b) all channel inputs are required to use the same codebook.

For a given $t$-user strategy $V,$ the \emph{maximum} 
probability of error is given by
\begin{equation}
\label{eqn:cap-maxpe}e_{\max}(F,\Phi,V)= \max_{u_{1},\dots,u_{t} \in\cM}
e(\{u_{1},\dots,u_{t}\},F,\Phi,V).
\end{equation}
Note that here the users $u_{1},\dots,u_{t}$ are not
necessarily distinct. It is straightforward to see that the $t$-fingerprinting
condition (\ref{eqn:cap-fingdef}) can now be expressed as
\begin{equation}
\label{cap-eqn:maxpe}e_{\max}(F,\Phi,V) \leq\varepsilon\text{ for every } V
\in\cV_{t}.
\end{equation}

\begin{definition}
\label{def:cap-capall} For $0 < \veps < 1,$ a number $R \ge0$ is an
$\veps$-\emph{achievable rate} for $q$-ary $t$-fingerprinting if for 
every $\delta>0$ and every $n$ sufficiently large, there exists a 
randomized $q$-ary code $(F,\Phi)$ of length $n$ with rate
\[
\frac{1}{n} \log_{q} M > R- \delta
\]
and maximum probability of error satisfying (\ref{cap-eqn:maxpe}).

The $\veps$-\emph{capacity} of $q$-ary $t$-fingerprinting 
$C_{t,q}(\veps)$ is the supremum of all such $\veps$-achievable rates. 
The \emph{capacity} of $q$-ary $t$-fingerprinting is the infimum of 
the $\veps$-capacities for $\veps>0$, i.e.,
\[
C_{t,q}=\lim_{\veps\rightarrow0}C_{t,q}(\veps).
\]
To proceed with the capacity $C_{t,q},$ we wish to consider coalitions of size
exactly $t.$ First, given any $t$-user strategy $V,$\ define the maximum
probability of error corresponding to \emph{coalitions of size $t$ alone\/}
as
\begin{equation}
\tilde{e}_{\max}(F,\Phi,V)=\max_{U:|U|=t}e(U,F,\Phi,V).\label{eqn:cap-maxpe2}%
\end{equation}
The capacity value $\tilde{C}_{t,q}$ corresponding to the above criterion
can be similarly defined.
\end{definition}

\begin{proposition}
\label{prop:cap-prop}%
\[
C_{t,q}=\tilde{C}_{t,q}.
\]
\end{proposition}

Clearly, $C_{t,q} \leq \tilde{C}_{t,q}.$ The proof of the opposite 
inequality is also almost obvious because any coalition of
$t$ pirates can simply ignore any subset of $t-\tau$ pirates when
devising a forged fingerpint $\bsy.$ A formal
version of this argument is provided by Lemma A.1 in the Appendix.
\remove{The proof of this statement is almost obvious. Clearly,
$\ C_{t,q}\leq\tilde{C}_{t,q}.$ Next, note that $\tilde{C}_{t,q}%
\leq\tilde{C}_{\tau,q}$ for any $\tau<t.$ Indeed,  any
coalition of $t$ pirates can simply ignore any subset of $t-\tau$
pirates when forging a new fingerprint $\bsy.$ In this case, the
distributor may retrieve any discarded pirate only with a risk of
picking up an innocent user, as the former becomes (equivalent to)
the latter. Thus, the $t$-pirate case always includes all possible
forgeries $V$ and optimal decoding strategies $\Phi$ that can be
executed on the coalitions of smaller size $\tau$. Therefore,
$\tilde{C}_{t,q}\leq\tilde{C}_{\tau,q}$ and the overall
capacity $C_{t,q}$ is always achieved on coalitions of size $t.$

We also refer to Lemma A.1 in the Appendix which compares $t$-coalitions
and $\tau$-coalitions, $\tau<t$, from a slightly more quantitative perspective.}

Similarly to the above, let us consider the 
{\em average} error probability 
\begin{equation}\label{eqn:cap-avgpe}
e_{\text{avg}}(F,\Phi,V)= \frac{1}{M^t}\sum_{u_1,\dots,u_t \in \cM}
e(\{u_1,\dots,u_t\},F,\Phi,V).
\end{equation}
(this quantity will be used in the derivation of upper bounds on the
capacity $C_{t,q}$) and the probability
\begin{equation}
\label{eqn:cap-avgpe2}\tilde{e}_{\text{avg}}(F,\Phi,V)= \frac{1}{\binom{M}{t}}
\sum_{U:|U|= t} e(U,F,\Phi,V).
\end{equation}
for coalitions of size exactly $t$. 
Define the capacity 
$\tilde{C}_{t,q}^{a}$ with respect to the latter error probability.

We make a remark on the relation between the average and maximum error
criteria. In general, it is true that the average error criterion yields a
higher capacity value compared to the maximum one. However, when randomization
is allowed, it is well-known for single-user channels that the capacity value
is the same for both the maximum and average error probability criteria (cf.
e.g., \cite[p.223, Prob. 5]{csis-kor}). We now extend this argument to the
current context of multi-user channels and fingerprinting to show that both
(\ref{eqn:cap-maxpe2}) and (\ref{eqn:cap-avgpe2}) lead to the same capacity value.

\begin{proposition}
\label{prop:cap-rccap}
\[
\tilde{C}_{t,q}=\tilde{C}_{t,q}^{a}.
\]
\end{proposition}

\medskip A formal proof is available in the Appendix. It follows because 
here we simply use a randomized code $(F,\Phi)$, which also includes all $M!$
permutations of any specific realization of  $(F,\Phi)$. Because of the
symmetry introduced by this, the error probability $e(U,F,\Phi,V)$ is the
same for all coalitions for a given $V$, and hence the average and
the maximum probability are the same.
\remove{In turn, all
coalitions $U$ then run through the same set of keys as code $(F,\Phi)$
runs through its  realizations. This also averages the error rates
$e(U,F,\Phi,V)$ for all coalitions $U.$}

\section{Lower bounds for binary $t$-secure codes, $t=2,3$}

\label{sect:cap-lb2}

In this section, we construct fingerprinting codes for $t=2,3,$ with error
probability decaying exponentially in $n$ and with higher rate than previous
constructions. The improvement is obtained by tailoring the decoder for the
\emph{typical} coalitions, i.e., the coalitions that occur with high
probability. We will say that an event occurs with high probability if the
probability that it fails is at most $\exp(-cn),$ where $c$ is a positive constant.

Our aim is to construct a sequence of randomized codes $(F_{n},\Phi
_{n}),n=1,2,\dots,$ with error probability
\[
\max_{V\in\cV_{t,n}}e_{\max}(F_{n},\Phi_{n},V)
\]
decaying to zero. By Proposition  \ref{prop:cap-prop}, it suffices to consider
only coalitions of size exactly $t$. Suppose, for every $n,$ there exists a
set $\cT_{t,n}\subseteq(\cQ^{n})^{t}$ such that for any coalition $U$ of size
$t,$ the observed fingerprints $f_{K,n}(U)$ belong to $\cT_{t,n}$ with high
probability. We will refer to a set with this property as a \emph{typical}
set. Thus, in constructing the required code it suffices to study 
the conditions that allow us to obtain vanishing
probability
\[
\Pr\left\{  \phi_{K,n}(\bsy)\notin U\big|f_{K,n}(U)=(\bsx_{1},\dots
,\bsx_{t})\right\}
\]
for any coalition $U$ of size $t,$ any typical $t$-tuple $(\bsx_{1}%
,\dots,\bsx_{t})$ of observed fingerprints, and any forgery $\bsy\in
\cE(\bsx_{1},\dots,\bsx_{t})$ as $n\rightarrow\infty.$ Our first result is a
lower bound on the fingerprinting capacity with 2 pirates over the binary
alphabet. \medskip

\begin{theorem}
\label{thm:cap-lb-2}
\[
C_{2,2}\ge\nicefrac14.
\]

\end{theorem}

\medskip

\begin{proof}
Fix $\cQ=\{0,1\}.$ Suppose that the encoding mapping $F$ assigns $M=2^{nR}$
fingerprints to the users choosing them uniformly and independently from all
$2^{n}$ different vectors. For $R<1/2,$ the fingerprints will be distinct with
high probability.

Given a small $\varepsilon>0,$ we define the typical set as the set of vector
pairs which agree in $l$ positions, where
\[
l\in I_{\varepsilon}=\left[  n\left(  \nicefrac{1}{2}-\varepsilon\right)
,n\left(  \nicefrac{1}{2}+\varepsilon\right)  \right]  .
\]
Consider any two pirates $u_{1}$ and $u_{2}.$ Notice that their observed
fingerprints form a typical pair  $(\bsx_{1},\bsx_{2})$ with high probability
for an arbitrarily small $\varepsilon.$ Hence,
$(\bsx_{1},\bsx_{2})$ agree in $l\in I_{\veps}$ positions. To create a forged
fingerprint $\bsy$ the pirates must fill the remaining $n-l$ positions. Let
$d_{1}=\qopname\relax{no}{dist}(\bsy,\bsx_{1})$ and $d_{2}%
=\qopname\relax{no}{dist}(\bsy,\bsx_{2}).$ Then $n-l\in I_{\varepsilon}$ and
therefore
\begin{equation}
d_{1}+d_{2}\in I_{\varepsilon}.\label{eqn:cap-lb2}%
\end{equation}
Given a forgery $\bsy,$ the decoder only considers typical pairs
$(\bsx_{1},\bsx_{2})$ from the codebook. Namely, the decoder takes any pair of
distances $(d_{1},d_{2})$ that satisfy (\ref{eqn:cap-lb2}) and constructs the
full lists $S_{\bsy}(d_{1})$ and $S_{\bsy}(d_{2})$ of the fingerprints located
at distances $d_{1}$ and $d_{2}$ from $\bsy.$ Each pair $(\bsx_{1}%
,\bsx_{2})\in S_{\bsy}(d_{1})\times S_{\bsy}(d_{2})$ is then discarded if they
simultaneously disagree with $\bsy$ in any position $s,$ i.e., $x_{1s}%
=x_{2s}\neq y_{s}.$ All  remaining pairs contain $\bsy$ in their envelope. For
each such pair  $(\bsx_{1},\bsx_{2}),$  the decoding is completed by choosing
the pirate $u_{i}$ whose fingerprint $\bsx_{i}$ has a smaller distance
$d_{i}=\qopname\relax{no}{dist}(\bsy,\bsx_{i}).$ Either user is chosen if
$d_{1}=d_{2}$.

Obviously, the fingerprints $\bsx_{1}$ and $\bsx_{2}$ that belong to the
factual pirates will not be discarded by the above decoding algorithm. The
following probabilistic analysis shows that for two innocent users, the
decoder discards their observed fingerprints $(\bsz_{1},\bsz_{2})$ with high
probability if the code rate
\[
R<\nicefrac{1}{4}.
\]
Indeed, for $(\bsz_{1},\bsz_{2})$ to be typical, they should agree in $l\in
I_{\veps}$ positions. In all these positions, $\bsz_{1},\bsz_{2}$ should also
agree with $\bsy$ to fulfill the marking assumption. In each of the remaining
$n-l$ positions, the vectors $\bsz_{1},\bsz_{2}$ are represented by only two
combinations, $(01)$ or $(10).$ The probability of choosing such a pair
$(\bsz_{1},\bsz_{2})$ in our random code equals
\[
P_{l}=\binom{n}{l}2^{n-l}/2^{2n}%
\]
and has exponential order of $2^{-n/2}$ for any $l\in I_{\varepsilon}.$
Furthermore, by the union bound, the total probability of choosing such a pair
in a random code of size $M=2^{nR}$ is at most
\[
\binom{M}{2}\sum_{l\in I_{\varepsilon}}\binom{n}{l}2^{n-l}/2^{2n}.
\]
This probability tends to $0$ exponentially fast for any rate $R<0.25.$

Similarly, consider a coalition $(\bsx_{1},\bsz_{2})$ that includes the
fingerprint $\bsx_{1}$ of an actual pirate and the fingerprint $\bsz_{2}$ of
an innocent user. Recall that $\bsx_{1}$ disagrees with $\bsy$ in $d_{1}$
positions. Then to be output instead of $\bsx_{1},$ the fingerprint $\bsz_{2}$
must agree with $\bsy$ in these positions and disagree with it in another set
of $d_{2}\leq d_{1}$ positions. The total probability of choosing such a
fingerprint $\bsz_{2}$ is at most
\[
M2^{n-d_{1}}/2^{n}.
\]
Since $d_{1}+d_{2}\in I_{\varepsilon}$ and $d_{2}\leq d_{1}$, we have
restriction $d_{2}\leq\nicefrac{n}{2}\left(  \nicefrac{1}{2}+\varepsilon
\right)  .$ In this case, the above probability tends to $0$ exponentially
fast for any rate $R<0.25.$ Thus, at least one pirate will be chosen from each
coalition, and  with high probability, no  remaining (innocent) users will be
chosen as pirates.
\end{proof}

\medskip

We note that considering typical coalitions enables us to improve the lower
bound $C_{2,2}\ge0.2075$ obtained in \cite{kaba}.

Next we establish a lower bound on the fingerprinting capacity with 3 pirates
over the binary alphabet. \medskip

\begin{theorem}
\label{thm:cap-lb-3}
\[
C_{3,2}\geq\nicefrac1{12}.
\]

\end{theorem}

\medskip

\begin{proof}
Suppose again that the encoding mapping $F$ assigns $M=2^{nR}$ fingerprints to
the users choosing them uniformly and independently from all $2^{n}$ different
vectors. For a triple $(\bsx_{1},\bsx_{2},\bsx_{3}),$ let
\[
\cL=\{s\in\lbrack n]:\;x_{1s}=x_{2s}=x_{3s}\},
\]%
\[
\cL_{ij}=\{s\in\lbrack n]:\;x_{is}=x_{js}\},\hspace{0.5em}i,j=1,2,3,\hspace
{0.5em}i\neq j,
\]
and let $l=|\cL|$, $l_{ij}=|\cL_{ij}|.$ Given a small $\veps>0,$ we say that
$(\bsx_{1},\bsx_{2},\bsx_{3})$ form a typical triple if
\begin{equation}
l\in J_{\veps}=[n(\nicefrac14-\veps),n(\nicefrac14+\veps)],\label{eqn:cap-lb3}%
\end{equation}%
\begin{equation}
l_{12},l_{13},l_{23}\in I_{\veps}%
=[n(\nicefrac12-\veps),n(\nicefrac12+\veps)].\label{eqn:cap-lb4}%
\end{equation}
For any three users $u_{1},u_{2},u_{3},$ note that the observed fingerprints
form a typical triple with high probability.

Using the same idea as before, we now take the observed fingerprints
$(\bsx_{1},\bsx_{2},\bsx_{3})$ to be a typical triple. A forged fingerprint
$\bsy$ agrees with all the three fingerprints on $\cL$ and takes arbitrarily
values $\{0,1\}$ on the remaining subset $[n]\backslash\cL $ positions. Let
$d_{i}=\qopname\relax{no}{dist}(\bsy,\bsx_{i})$ for $i=1,2,3.$ Note that every
position in $[n]\backslash\cL$ contributes 1 or 2 to the sum $d_{1}%
+d_{2}+d_{3}$ implying
\begin{equation}
\label{eqn:cap-lb5}n(\nicefrac34-\veps)\leq d_{1}+d_{2}+d_{3}\leq
n(\nicefrac32+2\veps).
\end{equation}

Given a forged fingerprint $\bsy,$ the decoder considers only typical triples
$(\bsx_{1},\bsx_{2},\bsx_{3})$ from the codebook. Each triple is then
discarded if the fingerprints in it simultaneously disagree with $\bsy$ in any
position $s,$ i.e., $x_{1s}=x_{2s}=x_{3s}\neq y_{s}.$ If a triple
$(\bsx_{1},\bsx_{2},\bsx_{3})$ is left, decoding is completed by choosing the
pirate whose fingerprint has the smallest distance to $\bsy$ among
$\bsx_{1},\bsx_{2},\bsx_{3}$.

Obviously, the fingerprints $(\bsx_{1},\bsx_{2},\bsx_{3})$ corresponding to
the factual pirates will not be discarded by the decoder. The following
probabilistic analysis shows that a randomly chosen code of rate
\begin{equation}
R<\nicefrac{1}{12}\label{eqn:cap-lb6}%
\end{equation}
enables the decoder to discard with high probability all typical
triples $(\bsz_{1},\bsz_{2},\bsz_{3})$ of fingerprints formed by
three innocent users. Indeed, a typical triple can be identified
only if the fingerprints in it simultaneously agree with $\bsy$ in
some subset of $l\in J_{\varepsilon}$ positions. To simplify our
analysis in this case, we can even ignore the extra conditions
(\ref{eqn:cap-lb4}) in any of the remaining $n-l$ positions. Thus,
we  allow the vectors $(\bsz_{1},\bsz_{2},\bsz_{3})$ to take on
any combination of binary symbols $\{0,1\}$ different from all
zeros or all ones. Given 6 such combinations, any typical triple
is chosen with probability at most
\[
P_{l}\leq\binom{n}{l}6^{n-l}/2^{3n}.
\]
We further observe that the total probability of choosing such a triple in a
random code of size $M=2^{nR}$ equals
\[
\binom{M}{3}\sum_{l\in J_{\varepsilon}}\binom{n}{l}6^{n-l}/2^{3n}%
\]
and tends to $0$ exponentially fast for any rate $R<\nicefrac1{12}$. Now
consider a slightly more involved case when the decoder locates the pirate
coalition $(\bsx_{1},\bsx_{2},\bsx_{3})$ along with another coalition
$(\bsx_{1},\bsz_{2},\bsz_{3})$ that includes the fingerprint $\bsx_{1}$ of an
actual pirate and the fingerprints $\bsz_{2},\bsz_{3}$ of two innocent users.
In what follows, we prove that a random code of rate (\ref{eqn:cap-lb6})
satisfies at least one of the following two conditions:

\begin{itemize}
\item[(i)] The decoder chooses $\bsx_{1}$ in the coalition $(\bsx_{1}%
,\bsz_{2},\bsz_{3})$ with high probability.

\item[(ii)] The coalition $(\bsx_{1},\bsz_{2},\bsz_{3})$ has vanishing probability.
\end{itemize}

Recall that $d_{1}=\qopname\relax{no}{dist}(\bsy,\bsx_{1}).$ Then an innocent
user, $\bsz_{2}$ say, can be output by the decoder if
$\qopname\relax{no}{dist}(\bsy,\bsz_{2})\leq d_{1}$. The probability that any
such $\bsz_{2}$ is chosen among $M$ random codewords is obviously at most
\[
2^{-n}M\sum_{i=0}^{d_{1}}\binom{n}{i}.
\]
Given a code of rate (\ref{eqn:cap-lb6}), this probability vanishes if
$\nicefrac{d_1}{n}\leq0.33.$ Therefore, condition (i) above fails if
\begin{equation}
{d_{1}}/{n}>0.33.\label{eqn:cap-lb7}%
\end{equation}
Now let us consider condition (ii) given this restriction. Consider a typical
coalition $(\bsx_{1},\bsz_{2},\bsz_{3}).$ We have
\[
l=|\{s\in\lbrack n]:\;x_{1s}=z_{2s}=z_{3s}=y_{s}\}|,
\]%
\[
l^{\prime}=|\{s\in\lbrack n]:\;z_{2s}=z_{3s}\neq x_{1s}\}|
\]
Thus, the vectors $\bsz_{2},\bsz_{3}$ have fixed values on the one subset  of
size $l,$ where these vectors  are equal to $\bsx_{1}$,  and on the other
non-overlapping subset of size  $l^{\prime}$, where the vectors $\bsz_{2}%
,\bsz_{3}$ are equal to the binary complement of  $\bsx_{1}.$ According to
conditions (\ref{eqn:cap-lb3}) and (\ref{eqn:cap-lb4}), $l\in J_{\varepsilon}$
and
\[
l^{\prime}=l_{23}-l\in J_{2\varepsilon}.
\]
In the remaining $n-l-l^{\prime}$ positions we have
\[
(z_{2s},z_{3s})\in\{(10),(01)\}.
\]
Summarizing the above arguments, we conclude that the total probability of
choosing such vectors $\bsz_{2},\bsz_{3}$ in the random code is bounded above
as
\[
2^{-2n}\binom{M}{2}\sum_{l\in J_{\varepsilon}}\sum_{l^{\prime}\in
J_{2\varepsilon}}\binom{n-d_{1}}{l}\binom{n-l}{l^{\prime}}2^{n-l-l^{\prime}}.
\]
Straightforward verification shows that this quantity vanishes given
conditions (\ref{eqn:cap-lb3}), (\ref{eqn:cap-lb4}), and (\ref{eqn:cap-lb7})
for a code of rate $R<0.086.$ Thus a random code of smaller rate
(\ref{eqn:cap-lb6}) discards all mixed coalitions of the form $(\bsx_{1}%
,\bsz_{2},\bsz_{3})$ with high probability.

The last remaining case, of a mixed coalition $(\bsx_{1},\bsx_{2},\bsz_{3})$,
is analyzed in a similar fashion (the analysis is simpler than the one above
and will be omitted).
\end{proof}

\section{A weak converse upper bound}

\label{sect:cap-ub-w}

In finding upper bounds on fingerprinting capacity (here and also in Section
\ref{sect:cap-ub-s}), we restrict our attention to \emph{memoryless} coalition
strategies in order to make the problem tractable and to obtain single-letter
expressions. Any upper bound on the capacity thus obtained will be also
valid in the original problem.

Let $\cW_{t}$ denote the family of discrete memoryless channels (DMCs) $W:
\cQ \times\dots\times\cQ \to\cQ$ with $t$ inputs that satisfy the marking
assumption for a single letter, i.e.,
\begin{equation}
\label{eqn:cap-chset}\cW_{t}=\{W:W(y|x,\dots,x)=0 \text{ if } y \neq x,
\forall x,y \in Q\}.
\end{equation}
Note that the above definition corresponds to the wide-sense envelope
$\cE_{W}(\cdot)$ defined in (\ref{eqn:intro-wenv}). For the narrow-sense
envelope $\cE_{N}(\cdot)$ (\ref{eqn:intro-nenv}) and other variations of the
problem it is possible to define similar communication channels and study
their upper bounds on capacity.

Observe that $\cW_{t}$ is a convex and compact set. Let $s \in\cS_{t}$, called
the ``state'', be an index which identifies the particular $W \in\cW_{t}$.
Hence, we will often write $W(y|x_{1},\dots,x_{t};s)$ for channels in
$\cW_{t}$.

We model a coalition's strategy by a (discrete memoryless) arbitrarily varying
channel (AVC), i.e., the state of the channel can vary from symbol to symbol.
For a given state sequence $\bss \in\cS_{t}^{n}$, the channel is given by
\begin{equation}
W^{n}(\bsy|\bsx_{1},\dots,\bsx_{t};\bss)=\prod_{l=1}^{n}W(y_{l}|x_{1l}%
,\dots,x_{tl};s_{l}).
\end{equation}
We denote the family of such channels $W^{n}(\cdot|\cdot,\dots,\cdot;\bss):
\cQ^{n} \times\dots\times\cQ^{n} \to\cQ^{n}, \bss \in\cS_{t}^{n}$ by
$\cW_{t}^{n}.$

Since the state sequence $\bss$ completely identifies the channel, we will
use
%$e_{\max}(F,\Phi,\bss)$ and
$e_{\text{avg}}(F,\Phi,\bss)$ to denote the error probability in
(\ref{eqn:cap-avgpe}).
%Similar to Definition \ref{def:cap-capall}, one may now define the {\em capacity}
%of $t$-fingerprinting over the class of arbitrarily varying memoryless
%strategies $\cW_t^n.$ As this definition for capacity takes into account
%only memoryless coalition strategies, it is an upper bound on the overall
%capacity of fingerprinting $C_{t,q}$ that takes into account all admissible
%coalition strategies (cf. Definition \ref{def:cap-capall}).

\subsection{The general case}

\begin{theorem}
\label{thm:cap-ub-w} Let $(F,\Phi)$ be a $q$-ary $t$-fingerprinting code with
$\veps$-error ($0 < \veps <1$) of length $n,$ rate $R,$ and $|\cK|$ keys, such
that $\veps q^{nR} \ge2^{t}.$ Then
\[
R \le\frac{1}{1-2t\veps} \left(  \max_{P_{K X_{1} \dots X_{t}}} \min_{W
\in\cW_{t}} I(X_{1},\dots,X_{t};Y|K) + \xi_{n} \right)
\]
where $\xi_{n}=t \log_{q}2/n,$ $X_{1},\dots,X_{t}, Y$ are $q$-ary r.v.'s,
$P_{Y|X_{1} \dots X_{t}}=W,$ $K$ is a r.v. taking values over a set of
cardinality $|\cK|$ and satisfying the Markov chain $K \leftrightarrow
X_{1},\dots,X_{t} \leftrightarrow Y,$ and the maximization is over joint
distributions
\begin{equation}
\label{eqn:cap-ub-w3}%
\begin{array}
[c]{l}%
P_{K X_{1} \dots X_{t}} = P_{K} \times P_{X_{1}|K} \times\dots\times
P_{X_{t}|K}\\
\text{with } P_{X_{1}|K} = \dots= P_{X_{t}|K}.
\end{array}
\end{equation}

\end{theorem}

\begin{proof}
%We show an upper bound on the capacity of fingerprinting over arbitrarily
%varying memoryless strategies under the average error criterion
%which in turn forms an upper bound on the fingerprinting
%capacity $C_{t,q}.$

Let $\cK$ be a set of keys and let $\{(f_{k},\phi_{k}),k \in\cK\}$ be a family
of codes with probability distribution $\pi(k)$ over $\cK.$ Since $(F,\Phi)$
is $t$-fingerprinting with $\veps$-error, it satisfies
\begin{equation}
\label{eqn:cap-perr}e_{\text{avg}}(F,\Phi,\bss) \leq\veps \text{ for every }
\bss \in\cS_{t}^{n}.
\end{equation}
Let $U_{1},\dots,U_{t}$ be independent r.v.'s uniformly distributed over the
message set $\{1,\dots,q^{nR}\}$ and let $K$ be a r.v. independent of
$U_{1},\dots,U_{t},$ and with probability distribution $\pi(k)$ over $\cK$.
Also, let
\begin{equation}
\label{eqn:cap-x1x2}\bsX_{i} \triangleq f_{K}(U_{i}), \quad i=1,\dots,t.
\end{equation}
Fix some $\bss \in\cS_{t}^{n}$ and let $\bsY$ be such that $P_{\bsY|\bsX_{1}%
,\dots,\bsX_{t}}=W^{n}(\cdot|\cdot,\dots,\cdot;\bss)$. Then, we have
\begin{equation}
\label{eqn:cap-perr2}\Pr(\phi_{K}(\bsY) \notin\{U_{1},\dots,U_{t}\})
\leq\veps,
\end{equation}
which follows from (\ref{eqn:cap-perr}). We also have the following 
Markov chain
\begin{equation}
\label{eqn:cap-markov}U_{1},\dots,U_{t},K \leftrightarrow\bsX_{1}%
,\dots,\bsX_{t} \overset{W^{n}}{\longleftrightarrow} \bsY.
\end{equation}
Now,
\begin{equation}
\label{eqn:cap-ub-w1}I(U_{1},\dots,U_{t};\bsY|K) = tnR - H(U_{1},\dots
,U_{t}|\bsY,K),
\end{equation}
because $U_{1},\dots,U_{t}$ are independent and uniformly distributed over
$\cM.$ The second term in (\ref{eqn:cap-ub-w1}) can be bounded above as
follows. Define $E_{i} = 1(\phi_{K}(\bsY) \neq U_{i}),$ $i=1,\dots,t.$ Let
$p_{i}=\Pr(E_{i}=0,E_{j}=1, j=1,\dots,t, j\neq i),$ $i=1,\dots,t.$ Since
$\phi_{K}(\bsY),E_{1},\dots,E_{t}$ are known given $K, \bsY, U_{1},\dots
,U_{t},$
\begin{align}
&  H(U_{1},\dots,U_{t}|\bsY,K)\nonumber\\
&  = H(U_{1},\dots,U_{t},E_{1},\dots,E_{t}|\bsY,\phi_{K}(\bsY),K)\nonumber\\
&  \leq t\log_{q}2 +H(U_{1},\dots,U_{t}|\bsY,\phi_{K}(\bsY),K,E_{1}%
,\dots,E_{t})\label{eq:H}\\
&  \leq t \log_{q}2 + \veps tnR + 2^{t} q^{-nR} tnR\nonumber\\
&  \quad+ \sum_{i=1}^{t} p_{i} H(U_{1}^{t}\backslash U_{i}|U_{i},\bsY,K,
E_{i}=0,E_{j}=1, j\neq i)\label{eqn:cap-ub-w2}\\
&  \leq t \log_{q}2 + (\veps + 2^{t} q^{-nR})tnR + (t-1)nR.\nonumber
\end{align}
Equation (\ref{eq:H}) holds true because $E_{1},\dots,E_{t}$ are binary r.v.'s
and the term $2^{t} q^{-nR} tnR$ in (\ref{eqn:cap-ub-w2}) follows from the
fact that there are at most $2^{t}$ remaining terms and each can be bounded
above by $q^{-nR} tnR.$ Using this in (\ref{eqn:cap-ub-w1}), we obtain
\begin{equation}
\label{eqn:cap-ub-wmain}nR(1-(\veps + 2^{t} q^{-nR})t) \le I(U_{1},\dots
,U_{t};\bsY|K) + t \log_{q}2.
\end{equation}
We now use the premise that $\veps q^{nR} \ge2^{t},$ together with
(\ref{eqn:cap-markov}) and the memoryless property of the channel, which
results in
\begin{align*}
R  &  \leq\frac{1}{1-2t\veps} \left( \frac{1}{n}I(U_{1},\dots,U_{t};\bsY|K)
+\xi_{n} \right) \\
&  \leq\frac{1}{1-2t\veps} \left(  \frac{1}{n} I(\bsX_{1},\dots,\bsX_{t}%
;\bsY|K) +\xi_{n} \right) \\
&  \leq\frac{1}{1-2t\veps} \left(  \frac{1}{n} \sum_{l=1}^{n} I(X_{1l}%
,\dots,X_{tl};Y_{l}|K) +\xi_{n} \right) .
\end{align*}
Moreover, since the above bound applies for every $\bss \in\cS_{t}^{n},$ i.e.,
for every $W^{n} \in\cW_{t}^{n},$
\begin{align*}
R  &  \leq\frac{1}{1-2t\veps} \left(  \frac{1}{n} \min_{W^{n} \in\cW_{t}^{n}}
\sum_{l=1}^{n} I(X_{1l},\dots,X_{tl};Y_{l}|K) +\xi_{n} \right) \\
&  = \frac{1}{1-2t\veps} \left(  \frac{1}{n} \sum_{l=1}^{n} \min_{W \in
\cW_{t}} I(X_{1l},\dots,X_{tl};Y_{l}|K) + \xi_{n} \right) ,
\end{align*}
because the minimization is over channels whose state may vary over $\cW_{t}$
for every letter. Note that $\bsX_{1},\dots,\bsX_{t}$ are independent and
identically distributed (i.i.d.) given $K$ (by (\ref{eqn:cap-x1x2})).
Therefore, given $K,$ for every $l \in[n]$, $X_{1l},\dots,X_{tl}$ are i.i.d.
Hence,
\[
R \leq\frac{1}{1-2t\veps} \left(  \max_{P_{KX_{1}\dots X_{t}}} \min_{W
\in\cW_{t}} I(X_{1},\dots,X_{t};Y|K) +\xi_{n} \right) .
\]
where the maximization is over joint distributions satisfying
(\ref{eqn:cap-ub-w3}).
\end{proof}

\begin{corollary}
\label{cor:cap-ub-w}
\begin{equation}
\label{eqn:cap-ub-w}C_{t,q} \le\min_{W \in\cW_{t}} \max_{P_{X_{1}\dots X_{t}}}
I(X_{1},\dots,X_{t};Y),
\end{equation}
where $X_{1},\dots,X_{t}, Y$ are $q$-ary r.v.'s, $P_{Y|X_{1} \dots X_{t}}=W$
and the maximization is over joint distributions such that $X_{1},\dots,X_{t}$
are i.i.d.
\end{corollary}

\begin{proof}
As we prove only a min-max type result, it becomes sufficient to consider only
``fixed'' memoryless coalition strategies, i.e., strategies that remain fixed
at every symbol instead of varying arbitrarily. In the subsequent text,
$W^{n}$ will denote the $n$-letter extension of a DMC $W.$

Consider any sequence of $t$-fingerprinting codes $(F_{n},\Phi_{n}),
n=1,2,\dots$ of rate $R$ and error $\veps_{n},$ where $\veps_{n}$ approaches 0
as $n$ increases. Then
\begin{equation}
e_{\text{avg}}(F,\Phi,W^{n}) \leq\veps_{n} \text{ for every } W \in\cW_{t}.
\end{equation}
Fix some $W \in\cW_{t}.$ We find that (\ref{eqn:cap-ub-wmain}) holds for every
$n.$ Therefore by the arguments in Theorem \ref{thm:cap-ub-w}
\begin{equation}
\label{eqn:cap-ub-w7}R \leq\frac{1}{1-\veps_{n}^{\prime}} \left(  \frac{1}{n}
\sum_{l=1}^{n} I(X_{1l},\dots,X_{tl};Y_{l}|K) +\xi_{n} \right) ,
\end{equation}
where both $\veps_{n}^{\prime}=(\veps_{n} + 2^{t} q^{-nR})t$ and $\xi_{n}$ approach
0 as $n \to\infty.$ Considering the inner term, we note that
\[
\frac{1}{n} \sum_{l=1}^{n} I(X_{1l},\dots, X_{tl};Y_{l}|K) \le I(X_{1l^{\ast}%
},\dots, X_{tl^{\ast}};Y_{l^{\ast}}|K=k^{\ast}),
\]
where $l^{\ast}=l^{\ast}(W)$ and $k^{\ast}=k^{\ast}(W)$ are the coordinate and key which maximize the
mutual information. The term on the r.h.s. is a function of $(P_{X_{1l^{\ast}%
}\dots X_{tl^{\ast}}|K=k^{\ast}},W).$ For every $l \in[n],$ $X_{1l}%
,\dots,X_{tl}$ are i.i.d. when conditioned on $K.$ Therefore this term is at
most
\[
\max_{P_{X_{1}\dots X_{t}}} I(X_{1},\dots, X_{t};Y),
\]
where $X_{1},\dots,X_{t},Y$ are $q$-ary r.v.'s with $P_{Y|X_{1},\dots,X_{t}%
}=W,$ and the maximization is over i.i.d. r.v.'s. Finally, since
(\ref{eqn:cap-ub-w7}) is true for every $W \in\cW_{t},$ we obtain the stated
result by taking $n \to\infty.$
\end{proof}

\subsection{The binary case}

\label{sect:cap-ub-w-bin} 
\begin{figure}[ptb]
\centering \includegraphics[width=2.0in]{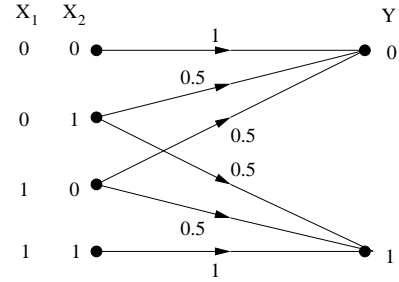}\caption{The uniform
channel with 2 pirates}%
\label{fig:cap-unifch}%
\end{figure}

Consider the case where $\cQ=\{0,1\}$. We would like
to evaluate the upper bound on $C_{t,2}$ given by Corollary \ref{cor:cap-ub-w}%
. Computing the exact optimum in this formula is a difficult problem. Instead
of attempting this, we will use a particular channel $W$ in
(\ref{eqn:cap-ub-w}) and compute a maximum on the prior distribution
$P_{X_{1}\dots X_{t}}$ for this channel. The resulting value of the rate gives
an upper bound on capacity $C_{t,2}.$ Let $W$ be the ``uniform channel''
defined by
\[
W(1|x_{1},\dots,x_{t})=\frac wt, \quad W(0|x_{1},\dots,x_{t})=1-\frac wt,
\]
where $w$ is the number of 1s among $x_{1},\dots,x_{t}.$ Fig.
\ref{fig:cap-unifch} shows the uniform channel for $t=2.$ Intuitively this
choice is the worst strategy of the coalition from the distributor's
perspective. 

If $X_{1},\dots,X_{t}$ are independent binary-valued r.v.'s with
$P(X_{i}=1)=p, 0 \leq p \leq1, i=1,\dots,t,$ and $Y$ is the output of the
uniform channel with inputs $X_{1},\dots,X_{t}$, we have $P(Y=1) = p$ and
\[
H(Y|X_{1},\dots,X_{t})= \sum_{i=0}^{t} \binom{t}{i} p^{i}(1-p)^{t-i} h\left(
\frac{i}{t}\right) .
\]
Evaluating the maximum mutual information in (\ref{eqn:cap-ub-w}) for this
channel gives a closed-form upper bound:

\begin{theorem}
\label{thm:cap-ub-w-bin}
\begin{align}
C_{t,2} & \leq\max_{p\in\lbrack0,1]}\Big\{h(p)-\sum_{i=0}^{t}\binom{t}{i}%
p^{i}(1-p)^{t-i}h\left(  \frac{i}{t}\right)  \Big\}\label{eqn:cap-ub-w-bin}\\
& \leq\frac{1}{t\ln2}.\label{eqn:cap-ub-w-bin-2}%
\end{align}

\end{theorem}

\medskip A proof of the estimate (\ref{eqn:cap-ub-w-bin-2}) is given in the Appendix.

\section{A strong converse upper bound}

\label{sect:cap-ub-s}

\subsection{The general case}

\begin{theorem}
\label{thm:cap-ub-s} For any $0 < \veps < 1,$
\begin{equation}
\label{eqn:cap-ub-s}C_{t,q}(\veps) \le\min_{W \in\cW_{t}} \max_{P_{X_{1}\dots
X_{t}}} \max_{i=1,\dots,t} I(X_{i};Y|X_{1}^{i-1},X_{i+1}^{t}),
\end{equation}
where $X_{1},\dots,X_{t}, Y$ are $q$-ary r.v.'s, $P_{Y|X_{1} \dots X_{t}}=W$
and the maximization is over joint distributions such that $X_{1},\dots,X_{t}$
are independent.
\end{theorem}

\begin{proof}
We borrow techniques from \cite{ahls-mac} in this proof. The result is proved
for the case $t=2$ to ease understanding. It is a straightforward extension
for arbitrary $t.$ In the proof, all logarithms are to the base $q.$

Consider a family of $n$-length codes $\{(f_{k},\phi_{k}),k \in\cK\}$ for $M$
users with probability distribution $\pi(k)$ over $\cK$ which is
$2$-fingerprinting with $\veps$-error ($0 < \veps < 1$). Therefore
\[
e_{\text{avg}}(F,\Phi,W^{n}) \leq\veps \text{ for every } W \in\cW_{2}.
\]
Let $\bsx_{i}^{(k)}=f_{k}(i)$ denote the fingerprints and $D_{i}^{(k)}=
\{\bsy: \phi_{k}(\bsy)=i\}$ denote the decoding regions for $i=1,\dots,M,$ $k
\in\cK.$ Then the above error criterion can be written as follows: For every
$W \in\cW_{2},$
\[
\sum_{k \in\cK} \pi(k) \frac{1}{M^{2}} \sum_{i,j=1}^{M} W^{n}(D_{i}^{(k)} \cup
D_{j}^{(k)} | \bsx_{i}^{(k)}, \bsx_{j}^{(k)}) \ge1-\veps.
\]
Fix some $W \in\cW_{2}.$ There exists a $k^{\ast}=k^{\ast}(W)\in\cK$ such that
\[
\frac{1}{M^{2}} \sum_{i,j=1}^{M} W^{n}(D_{i}^{(k^{\ast})} \cup D_{j}%
^{(k^{\ast})} | \bsx_{i}^{(k^{\ast})}, \bsx_{j}^{(k^{\ast})}) \ge1-\veps.
\]
Hereafter, we drop the superscript $k^{\ast}$ for simplicity. Consequently,
either
\begin{align}
&  \frac{1}{M^{2}} \sum_{i,j=1}^{M} W^{n}(D_{i} | \bsx_{i}, \bsx_{j}) \ge
\frac{1-\veps}{2}\label{eqn:cap-ub-s1}\\
\text{or} \quad &  \frac{1}{M^{2}} \sum_{i,j=1}^{M} W^{n}(D_{j} | \bsx_{i},
\bsx_{j}) \ge\frac{1-\veps}{2}\label{eqn:cap-ub-s2}%
\end{align}
must be true. Let us assume (\ref{eqn:cap-ub-s1}) is true. We first find a
subset $\cA$ of ``good'' pairs of users (messages) for $W.$ Define
\begin{equation}
\label{eqn:cap-ub-s-defA}\cA \triangleq\{(i,j):W^{n}(D_{i} | \bsx_{i},
\bsx_{j}) \ge1-\bar{\veps},1 \le i,j \le M\},
\end{equation}
where $\bar{\veps}$ is such that $0 < 1-\bar{\veps} < (1-\veps)/2.$ Then
\begin{equation}
\label{eqn:cap-ub-s3}|\cA| \ge(1-\veps^{\ast})M^{2}, \text{ where }
\veps^{\ast}\triangleq\frac{1+\veps}{2\bar{\veps}}.
\end{equation}
Next, we derive a subset $\bar{\cA}$ of the ``good'' pairs where approximate
independence holds between the fingerprints (codewords) corresponding to a
pair of users (messages) uniformly distributed over this subset. This is
needed to restrict the maximization in the final result (\ref{eqn:cap-ub-s})
to joint distributions where the r.v.'s are independent.
%\medskip

\begin{lemma}
\label{lem:cap-wring} \cite{ahls-mac} Let $\cC = \{\bsx_{1},\dots,\bsx_{M}\}
\subseteq\cQ^{n},$ $\cA \subset\{1,\dots,M\} \times\{1,\dots,M\}$ with $|\cA|
\ge(1-\veps^{\ast})M^{2},$ $0< \veps^{\ast}<1.$ Then for any $0 < \gamma<
\veps^{\ast}/(1-\veps^{\ast}),$ $0 \le\lambda<1,$ there exist $l_{1}%
,\dots,l_{r} \in[n],$ where $r \le\veps^{\ast}/(\gamma(1-\veps^{\ast})),$ and
some $(\bar{x}_{1},\bar{x}^{\prime}_{1}),\dots,(\bar{x}_{r},\bar{x}^{\prime
}_{r}),$ such that for $\bar{\cA} \triangleq\left\{ (i,j) \in\cA: x_{il_{m}%
}=\bar{x}_{m},x_{jl_{m}}=\bar{x}^{\prime}_{m}, \forall m \in[r] \right\} $

\begin{itemize}
\item[(a)] $\left| \bar{\cA}\right|  \ge\lambda^{r} \left| \cA\right| $, and

\item[(b)] For all $x_{1},x_{2} \in\cQ,$ $l \in[n],$
\begin{align*}
&  (1+\gamma)\Pr(\bar{X}_{1l}=x_{1})\Pr(\bar{X}_{2l}=x_{2}) - \gamma-
|\cQ|^{2}\lambda\\
&  \le\Pr(\bar{X}_{1l}=x_{1},\bar{X}_{2l}=x_{2})\\
&  \le\max\left\{ (1+\gamma)\Pr(\bar{X}_{1l}=x_{1})\Pr(\bar{X}_{2l}%
=x_{2}),\lambda\right\} ,
\end{align*}
where $(\bar{\bsX}_{1},\bar{\bsX}_{2})$ is a pair of r.v.'s with uniform
distribution on $\{(\bsx_{i},\bsx_{j}): (i,j) \in\bar{\cA}\}.$
\end{itemize}
\end{lemma}

\medskip

Applying Lemma \ref{lem:cap-wring} to $\cA$ as in (\ref{eqn:cap-ub-s-defA})
with parameters $\gamma= n^{-1/2},$ $\lambda=n^{-1},$ we obtain
\begin{equation}
\label{eqn:cap-ub-s8}\left| \bar{\cA}\right|  \ge\lambda^{r} \left| \cA\right|
, \text{ for some } r \le n^{1/2}\veps^{\ast}/(1-\veps^{\ast}).
\end{equation}

For $j=1,\dots,M,$ define $\cB(j)=\{i: (i,j) \in\bar{\cA}, 1 \le i \le M\}.$
Observe that the subcode corresponding to $\cB(j)$ is a ``good'' code for the
single-user channel obtained by fixing the second input to $j.$ Thus, the
single-user strong converse given below holds for this subcode.
%\medskip

\begin{lemma}
\label{lem:cap-aug} \cite{augustin} If $(f,\phi)$ is a code with codewords
$\{\bsx_{1},\dots,\bsx_{M}\} \subseteq\cQ^{n}$ and decoding regions $D_{i},
i=1,\dots, M,$ for the (non-stationary) single-user DMC $\{W_{l}%
\}_{l=1}^{\infty},$ such that for every $i=1,\dots,M,$ $\Pr(D_{i}|\bsx_{i})
\ge1-\bar{\veps},$ $0 < \bar{\veps} <1,$ then
\[
\log M \le\sum_{l=1}^{n} I(X_{l};Y_{l}) + O(n^{1/2}),
\]
where $\bsX$ is distributed uniformly on the set of codewords.
\end{lemma}

\medskip

Using Lemma \ref{lem:cap-aug} on the subcode $\cB(j),$
\begin{equation}
\label{eqn:cap-ub-s4}\log|\cB(j)| \le\sum_{l=1}^{n} I(\bar{X}_{1l};\bar{Y}%
_{l}|\bar{X}_{2l}=x_{jl}) + O(n^{1/2}),
\end{equation}
where $(\bar{\bsX}_{1},\bar{\bsX}_{2})$ are distributed as in Lemma
\ref{lem:cap-wring} and $P_{\bar{\bsY}|\bar{\bsX}_{1},\bar{\bsX}_{2}}=W^{n}.$
Furthermore, using (\ref{eqn:cap-ub-s4}), we obtain
\begin{align}
&  |\bar{\cA}|^{-1} \sum_{(i,j)\in\bar{\cA}} \log|\cB(j)|\nonumber\\
&  \le|\bar{\cA}|^{-1} \sum_{(i,j)\in\bar{\cA}} \sum_{l=1}^{n} I(\bar{X}%
_{1l};\bar{Y}_{l}|\bar{X}_{2l}=x_{jl}) \sum_{x \in\cQ} 1(x_{jl}=x)\nonumber\\
&  \hspace{17em} + O(n^{1/2})\nonumber\\
&  = \sum_{l=1}^{n} \sum_{x \in\cQ} |\bar{\cA}|^{-1} \sum_{(i,j)\in\bar{\cA}}
1(x_{jl}=x) I(\bar{X}_{1l};\bar{Y}_{l}|\bar{X}_{2l}=x_{jl})\nonumber\\
&  \hspace{17em} + O(n^{1/2})\nonumber\\
&  = \sum_{l=1}^{n} I(\bar{X}_{1l};\bar{Y}_{l}|\bar{X}_{2l}) + O(n^{1/2}%
),\label{eqn:cap-ub-s5}%
\end{align}
since $\Pr(\bar{X}_{2l}=x)=|\bar{\cA}|^{-1} \sum_{(i,j)\in\bar{\cA}}
1(x_{jl}=x)$ for $l \in[n].$ We next establish a lower bound on the left-side
term in order to obtain an inequality for $M.$
\begin{align}
&  |\bar{\cA}|^{-1} \sum_{(i,j)\in\bar{\cA}} \log|\cB(j)|= |\bar{\cA}|^{-1}
\sum_{j=1}^{M} |\cB(j)| \log|\cB(j)|\label{eqn:cap-ub-s6}\\
&  \ge|\bar{\cA}|^{-1} \sum_{j:|\cB(j)| \ge\frac{1-\veps^{\ast}}{n}M
\lambda^{r}} |\cB(j)| \log|\cB(j)|\nonumber\\
&  \ge|\bar{\cA}|^{-1} \log\left(  \frac{1-\veps^{\ast}}{n}M \lambda^{r}
\right)  \sum_{j:|\cB(j)| \ge\frac{1-\veps^{\ast}}{n}M \lambda^{r}}
|\cB(j)|\label{eqn:cap-ub-ss1}%
\end{align}
where (\ref{eqn:cap-ub-s6}) follows from the definition of $\cB(j).$ Now,
\begin{align*}
&  \sum_{j:|\cB(j)| \ge\frac{1-\veps^{\ast}}{n}M \lambda^{r}} |\cB(j)|\\
&  = \sum_{j=1}^{M} |\cB(j)| - \sum_{j:|\cB(j)| < \frac{1-\veps^{\ast}}{n}M
\lambda^{r}} |\cB(j)|\\
&  \geq|\bar{\cA}| - \frac{1-\veps^{\ast}}{n}M^{2} \lambda^{r}\\
&  \geq|\bar{\cA}| - \frac{1}{n}|\bar{\cA}|
\end{align*}
by using (\ref{eqn:cap-ub-s3}) and (\ref{eqn:cap-ub-s8}). Using this
inequality in (\ref{eqn:cap-ub-ss1}), we get
\begin{equation}
\label{eqn:cap-ub-s7}|\bar{\cA}|^{-1} \sum_{(i,j)\in\bar{\cA}} \log|\cB(j)|
\ge\left(  1-\frac{1}{n} \right)  \log\left( \frac{1-\veps^{\ast}}{n}M
\lambda^{r}\right) .
\end{equation}
Combining (\ref{eqn:cap-ub-s5}), (\ref{eqn:cap-ub-s7}) and
(\ref{eqn:cap-ub-s8}),
\begin{align}
\log M  &  \le\left( 1+\frac{1}{n-1}\right)  \left( \sum_{l=1}^{n} I(\bar
{X}_{1l};\bar{Y}_{l}|\bar{X}_{2l}) + O(n^{1/2}) \right) \nonumber\\
&  \qquad- \log(1-\veps^{\ast}) + \log n + \frac{\veps^{\ast}}{1-\veps^{\ast}%
}n^{1/2}\log n\nonumber\\
&  \le\sum_{l=1}^{n} I(\bar{X}_{1l};\bar{Y}_{l}|\bar{X}_{2l}) + O(n^{1/2} \log
n).\label{eqn:cap-ub-s9}%
\end{align}
Although the above inequality resembles what is needed in the theorem, note
that $\bar{X}_{1l}$ and $\bar{X}_{2l}$ are not necessarily independent. For $l
\in[n],$ let $(X_{1l},X_{2l},Y_{l})$ be r.v.'s with distribution
\begin{align*}
&  \Pr(X_{1l}=x_{1},X_{2l}=x_{2},Y_{l}=y) =\\
&  \qquad\Pr(\bar{X}_{1l}=x_{1})\Pr(\bar{X}_{2l}=x_{2})W(y|x_{1},x_{2})
\end{align*}
for all $x_{1},x_{2},y \in\cQ.$ From Lemma \ref{lem:cap-wring}(b), for
$n^{-1/2} \ge|\cQ|^{2} n^{-1}$ and every $l \in[n]$
\begin{align*}
&  (1+n^{-1/2})\Pr(\bar{X}_{1l}=x_{1})\Pr(\bar{X}_{2l}=x_{2}) - 2n^{-1/2}\\
&  \le\Pr(\bar{X}_{1l}=x_{1},\bar{X}_{2l}=x_{2})\\
&  \le(1+n^{-1/2})\Pr(\bar{X}_{1l}=x_{1})\Pr(\bar{X}_{2l}=x_{2})+n^{-1},
\end{align*}
\begin{align*}
\text{i.e., }  &  |\Pr(X_{1l}=x_{1},X_{2l}=x_{2})\\
&  \qquad-\Pr(\bar{X}_{1l}=x_{1},\bar{X}_{2l}=x_{2})| \le2n^{-1/2}.
\end{align*}
Thus, by the uniform continuity of mutual information, for all $l \in[n],$
\[
|I(X_{1l};Y_{l}|X_{2l})-I(\bar{X}_{1l};\bar{Y}_{l}|\bar{X}_{2l})| \le
\alpha_{n},
\]
where $\alpha_{n} \to0$ as $n \to\infty.$ Together with (\ref{eqn:cap-ub-s9})
and dividing by $n,$
\begin{equation}
\label{eqn:cap-ub-s11}R \le\max_{P_{X_{1}X_{2}}=P_{X_{1}}P_{X_{2}}}
I(X_{1};Y|X_{2}) + \beta_{n},
\end{equation}
where $\beta_{n} = \alpha_{n} + O(n^{-1/2}\log n) \to0$ as $n \to\infty.$
Similarly, assuming (\ref{eqn:cap-ub-s2}) is true, one can prove
\begin{equation}
\label{eqn:cap-ub-s12}R \le\max_{P_{X_{1}X_{2}}=P_{X_{1}}P_{X_{2}}}
I(X_{2};Y|X_{1}) + \beta^{\prime}_{n}.
\end{equation}
Since either (\ref{eqn:cap-ub-s11}) or (\ref{eqn:cap-ub-s12}) holds for every
$W \in\cW_{2},$ taking $n\to\infty$ concludes the proof.
\end{proof}

\subsection{The binary case}

\label{sect:cap-ub-s-bin} Fix $\cQ=\{0,1\}$. For the case of $t=2$ and $t=3,$
we again pick the uniform channel and obtain upper bounds on the expression in
Theorem \ref{thm:cap-ub-s}, which turn out to be stronger than the bounds
resulting from (\ref{eqn:cap-ub-w-bin}). The calculations become quite tedious
for larger $t.$ For $t=2,$ let $X_{1},X_{2}$ be independent binary-valued
r.v.'s with $P(X_{i}=1)=p_{i}, 0 \leq p_{i} \leq1, i=1,2,$ and let $Y$ be the
output of the uniform channel with inputs $X_{1}$ and $X_{2}.$ We have
\begin{align*}
H(Y|X_{2}) &  =(1-p_{2})h\left( \frac{p_{1}}{2}\right)  + p_{2}h\left(
\frac{1-p_{1}}{2}\right) \\
H(Y|X_{1},X_{2})  &  =(1-p_{1})p_{2} + p_{1}(1-p_{2}).
\end{align*}
Computing the maximum conditional mutual information gives $C_{2,2} \le0.322.$
A similar computation for $t=3$ yields
\begin{align*}
&  H(Y|X_{2},X_{3})\\
&  = (1-p_{2})(1-p_{3})h\left( \frac{p_{1}}{3}\right)  + (1-p_{2})p_{3}h\left(
\frac{1+p_{1}}{3}\right) \\
&  \quad+ p_{2}(1-p_{3})h\left( \frac{1+p_{1}}{3}\right)  + p_{2}p_{3}h\left(
\frac{1-p_{1}}{3}\right) ,\\
&  H(Y|X_{1},X_{2},X_{3})\\
&  = \left( 1-p_{1}p_{2}p_{3}-(1-p_{1})(1-p_{2})(1-p_{3})\right)  h\left(
\frac{1}{3}\right) ,
\end{align*}
and the maximization gives $C_{3,2} \le0.199.$ Combining these upper bounds
with our lower bounds from Theorem \ref{thm:cap-lb-2} and Theorem
\ref{thm:cap-lb-3} we obtain:
%the two main results of this paper.

\begin{theorem}
\label{thm:cap-2bin}
\[
0.25 \le C_{2,2} \le0.322.
\]
\[
0.083\leq C_{3,2} \leq0.199.
\]

\end{theorem}

\section{Conclusion}

\label{sect:cap-concl}

In this paper, we prove new lower bounds on the maximum rate of binary
fingerprinting codes for 2 and 3 pirates by considering typical coalitions
which improves the random coding results obtained previously in the
literature. We also prove several new upper bounds on fingerprinting capacity
relying upon converse theorems for a class of channels which are similar to
the multiple-access channel. Our results establish for the binary case,
$C_{t,2}\leq(t\ln2)^{-1}.$ Combined with the result of \cite{tardos} this
implies that $O(1/t^{2})\leq C_{t,2}\leq O(1/t).$ For the general case with
arbitrary alphabets, we have established some upper bounds on the capacity
involving single-letter mutual information quantities.

\appendix

\subsection{A lemma on the size of coalitions}

{\em Lemma A.1:}
Let $(F,\Phi)$ be a randomized code of size at least $2t-1.$
Assume that 
  \begin{equation}\label{eq:assumption}
\tilde{e}_{\max}(F,\Phi,V) \le \veps \text{ for every } V \in \cV_t.
   \end{equation}
Then for any $\tau \le t,$ 
$$\tilde{e}_{\max}(F,\Phi,V) \le 2 \veps \text{ for every } V \in \cV_\tau.$$

\begin{proof}
For simplicity of presentation we take $\tau=t-1.$ The general
case of $1\le \tau<t$ can be established with only minor changes to the
proof below.
For any $V \in \cV_{t-1},$ let us define a $V^\prime \in \cV_t$ where
\begin{align*}
V^\prime(\bsy|\bsx_1,\dots,\bsx_{t-1},\bsx_t) &= 
V(\bsy|\bsx_1,\dots,\bsx_{t-1}), \\ 
&\forall \bsx_1,\dots,\bsx_t,\bsy \in \cQ^n. 
\end{align*}
Then, for any coalition $U$ of size $t-1,$ and any user $u \notin U,$
\begin{align}
& e(U, F, \Phi, V) \nonumber\\
& = {\sf E}_{K} \sum_{\substack{ \bsy: 
\\ \phi_K(\bsy) \notin U}} V(\bsy|f_K(U)) \nonumber \\
& = {\sf E}_{K}\sum_{\substack{ \bsy: 
\\ \phi_K(\bsy) \notin U}}  V^\prime(\bsy|f_K(U),f_K(u)) \nonumber\\
& = {\sf E}_{K} \Big[ \sum_{\substack{ \bsy: 
\\ \phi_K(\bsy) \notin U'}} V^\prime(\bsy|f_K(U^\prime))
 + \sum_{\substack{ \bsy: 
\\ \phi_K(\bsy) = u}} V^\prime(\bsy|f_K(U^\prime)) \Big], \label{eq:n}
\end{align}
where $U^\prime=U \cup \{u\}.$ The first term in the last equation satisfies 
\begin{equation}\label{eq:l}
e(U^\prime, F, \Phi, V^\prime) \le \veps
\end{equation}
by the assumption of the lemma. 
We will next show that the second term in (\ref{eq:n}) 
is also at most $\veps.$ Suppose for the sake of 
contradiction that
$$ {\sf E}_{K} \sum_{\substack{ \bsy:\\ 
\phi_K(\bsy) = u}} V^\prime(\bsy|f_K(U^\prime)) > \veps.$$ 
Let $u^\prime \notin U^\prime$ and $U^{\prime \prime}= U \cup \{u^\prime \}$ 
(we assume that the size of the code is at least $t+2,$ or at least $2t-1$ 
in the general case).
Then
\begin{align*}
& e(U^{\prime \prime}, F, \Phi, V^\prime) \\
& = {\sf E}_{K} \sum_{\substack{ \bsy: 
\\ \phi_K(\bsy) \notin U^{\prime \prime}}} 
V^\prime(\bsy|f_K(U^{\prime \prime}))\\
& \geq {\sf E}_{K} \sum_{\substack{ \bsy: 
\phi_K(\bsy)=u}} V^\prime(\bsy|f_K(U^\prime)) > \veps.
\end{align*}
But this contradicts our initial assumption (\ref{eq:assumption}).\end{proof}

\subsection{Proof of Proposition \ref{prop:cap-rccap}}

It is clear that $\tilde{C}_{t,q} \le\tilde{C}_{t,q}^{a}.$ Therefore, it
is enough to show that for every randomized code $(F,\Phi)$, there exists
another randomized code $(F^{\ast},\Phi^{\ast})$ of the same rate such that
$\tilde{e}_{\max}(F^{\ast},\Phi^{\ast},V)=\tilde{e}_{\text{avg}}(F,\Phi,V)$
for every channel $V.$

We are given $\{(f_{k},\phi_{k}), k \in\cK\}.$ Let $\sigma\in\Sigma$ identify
a particular permutation from the set of all permutations on the message set
$\cM.$ Choose $\sigma$ uniformly at random from $\Sigma$ and construct a new
key ${\kappa} \triangleq(k,\sigma).$ Define
\[
f^{\ast}_{{\kappa}}(\cdot) \triangleq f_{k}(\sigma(\cdot)),
\]
\[
\phi^{\ast}_{{\kappa}}(\cdot) \triangleq\sigma^{-1}(\phi_{k}(\cdot)).
\]
Let $(F^{\ast},\Phi^{\ast})$ be the randomized code corresponding to the
family $\{(f^{\ast}_{{\kappa}},\phi^{\ast}_{{\kappa}}), {\kappa} \in
\cK \times\Sigma\}.$ Then, for every channel $V,$ $\tilde{e}_{\text{avg}%
}(F^{\ast},\Phi^{\ast},V)= \tilde{e}_{\text{avg}}(F,\Phi,V)$.
Furthermore, for any $U \subseteq\cM,$ $|U|=t,$
\begin{align*}
&  e(U,F^{\ast},\Phi^{\ast},V)\\
&  = \frac{1}{M!} \sum_{\sigma\in\Sigma} \sum_{k \in\cK} \pi(k)
\sum _{\substack{\bsy: \\\phi_{k}(\bsy) \notin\sigma(U)}}
V(\bsy|f_{k}(\sigma(U)))
\end{align*}
which does not depend on the subset $U$ because of the averaging over all
permutations. This implies $\tilde{e}_{\max}(F^{\ast},\Phi^{\ast},V)=
\tilde{e}_{\text{avg}}(F^{\ast},\Phi^{\ast},V).$

\subsection{Proof of Theorem \ref{thm:cap-ub-w-bin}}

Our goal is to estimate $\max_{p\in\lbrack0,1]}u(p,t),$ where we use the
following notation%
\[
u(p,t)=h(p)-\sum_{i=0}^{t}\alpha_{i}h\left(  \ts\frac{i}{t}\right)  .
\]%
\[
\alpha_{i}=\binom{t}{i}p^{i}(1-p)^{t-i}.
\]
First, note that $h$ is a concave function, and therefore $u(p,t)$ is
non-negative for all $p\in\lbrack0,1].$ Bernstein proved that the sequence of
polynomials $B_{t}(p)=\sum_{i=0}^{t}\alpha_{i}f(i/t),t=1,2,\dots,$ where $f$
is a function continuous on $[0,1]$, provides a uniform approximation to $f$
on $[0,1].$ His proof, found for instance in Feller \cite{fel71} \S 7.2,
relies on the weak law of large numbers.
Refining the proof in the case of the function $h$, we show that for any
$p\in\lbrack0,1]$ and any $t,$
\begin{equation}
u(p,t)\leq\frac{1}{t\ln2}.\label{ld}%
\end{equation}
It suffices to consider the case $p\in(0,1/2].$ \ Given some $x=i/t,$ let us
write a quadratic Taylor approximation for $h(x):$
\begin{equation}
h\left(  x\right)  =h(p)+(x-p)\log_{2}\frac{1-p}{p}+\frac{(x-p)^{2}}%
{2}a(x)\label{ld0}%
\end{equation}
where the coefficient $a(x)$ depends on $x,$ since $a(x)=h^{\prime\prime
}(\gamma)$ \ for some $\gamma\in\lbrack x,p].$ We shall also consider the
residual function
\[
\tilde{g}\left(  x\right)  =h\left(  x\right)  -h(p)-(x-p)\log_{2}\frac
{1-p}{p}.
\]
The main part of our proof is to show that for any $x\in\lbrack0,1],$
\begin{equation}
2p^{-2}\log_{2}\left(  1-p\right)  \leq a(x)\leq0.\label{ld2}%
\end{equation}
\ The right inequality is obvious since $h^{\prime\prime}\left(  x\right)  <0$
for all $0<x<1.$ \ The left inequality will be proven in two steps. \

Let us take any point $x_{0}\in\lbrack0,p].$ Then we compare $\tilde{g}\left(
x\right)  $ with the quadratic function
\[
g_{x_{0}}(x)=a(x_{0})\frac{(x-p)^{2}}{2}%
\]
on the entire interval $x\in\lbrack0,p].$ We first prove that functions
$g_{x_{0}}(x)$ and $\tilde{g}\left(  x\right)  $ coincide at only two points,
namely $p$ and $x_{0}.$ Indeed, let us assume that there exists a third such
point $x_{1}.$ Without loss of generality, let $x_{0}<x_{1}<p.$ \ The
functions $g_{x_{0}}(x)$ and $\tilde{g}\left(  x\right)  $ coincide at the
ends of both intervals $[x_{0},x_{1}]$ and $[x_{1},p]$; therefore there exist
two points $\theta^{\prime}\in\left(  x_{0},x_{1}\right)  $ and $\theta
^{\prime\prime}\in\left(  x_{1},p\right)  $ where both functions have equal
derivatives:
\begin{align*}
a(x_{0})(\theta^{\prime}-p)  & =\log_{2}\frac{1-\theta^{\prime}}%
{\theta^{\prime}}-\log_{2}\frac{1-p}{p};\\
a(x_{0})(\theta^{\prime\prime}-p)  & =\log_{2}\frac{1-\theta^{\prime\prime}%
}{\theta^{\prime\prime}}-\log_{2}\frac{1-p}{p}.
\end{align*}
The left sides of both equalities represent a linear function of $\theta$
given by $a(x_{0})(\theta-p)$ whereas the right sides represent a convex
function $\log_{2}\frac{1-\theta}{\theta}-\log_{2}\frac{1-p}{p}.$ A linear
function can intersect a convex function at no more than two points. This
leads to a contradiction, which shows that $x_{0}=x_{1}$ and that the
functions $g_{x_{0}}(x)$ and $\tilde{g}\left(  x\right)  $ intersect at two
points $p$ and $x_{0}.$

\begin{figure}[ptb]
\centering \includegraphics[height=2.0in]{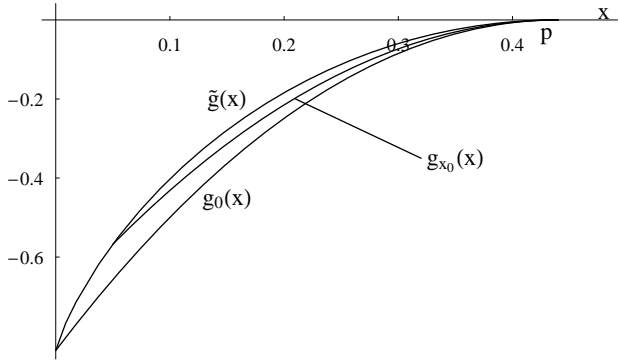}\caption{To the proof that
$a(x_{0})>a(0).$}%
\label{fig:parabolas}%
\end{figure}

Our next step is to find the minimum $a(x)\leq0$ for all $x\in\lbrack0,p]. $
\ Compare the function $g_{x_{0}}(x)$ with $g_{0}(x)$ for any parameter
$x_{0}\in(0,p].$ Now we use the fact that both functions intersect $\tilde
{g}\left(  x\right)  $ at only two points, one of which is $x=p.$ However,
$g_{0}(x)$ has its second intersection $x=0$ to the left of $x_{0}.$ Thus,
$g_{0}(x)<g_{x_{0}}(x)$ for $0\le x<p$ and therefore, $a(x_{0})>a(0)$ (see
Fig.~\ref{fig:parabolas}). Now we conclude that
\[
a(0)=\min_{x\in\lbrack0,p]}a(x).
\]

Finally, we find $a(0)$ using the equality $g_{0}(0)=\tilde{g}\left(  0\right)
,$ which gives $a(0)=2p^{-2}\log_{2}(1-p).$

The second interval $x\in\lbrack p,1]$ can be considered similarly$.$ Again,
we use the same arguments and conclude that the end point $x=1$ gives the
minimum $a(1)=\min_{x\in\lbrack p,1]}a(x).$ Direct calculation also shows that
the global minimum is achieved at $0$ as $a(0)<a(1)$ for all $p<1/2,$ and
$a(1)=a(0)$ for $p=1/2.$ This gives us the left inequality in (\ref{ld2}) and
shows that for any $p\leq1/2$ and any $x\in[0,1],$
\[
h\left(  x\right)  \geq h(p)+(x-p)\log_{2}\frac{1-p}{p}+(x-p)^{2}\frac
{\log_{2}\left(  1-p\right)  }{p^{2}}. \label{ld1}
\]
Let us take $x=i/t, i=0,1,\dots,t$ and substitute the above estimate into the
expression for $u(p,t).$ In this substitution, we also use the first two
moments of the binomial distribution $\left\{  \alpha_{i}\right\} ,$ which
gives
\[
S_{1}\triangleq\sum_{i=0}^{t}\alpha_{i}\left(  \ts\frac{i}{t}-p\right)  =0
\]
\[
S_{2} \triangleq\sum_{i=0}^{t}\alpha_{i}\left(  \ts\frac{i}{t}-p\right)
^{2}=\frac{p(1-p)}{t}.
\]
Then
\begin{align*}
u(p,t) & \leq-\log_{2}\frac{1-p}{p}S_{1}-\frac{\log_{2}(1-p)}{p^{2}}S_{2}\\
&  \leq-\frac{(1-p)\ln(1-p)}{pt\ln2}.
\end{align*}
Finally, it is easy to verify that the function $-\frac{(1-p)\ln(1-p)}{p}$
monotonically decreases on the interval $[0,\frac{1}{2}]$ and achieves its
maximum $1$ at $p=0.$ This establishes (\ref{ld}) and hence the bound
(\ref{eqn:cap-ub-w-bin-2}).

\section*{Acknowledgments}

The authors would like to thank Prakash Narayan and G\'{a}bor Tardos for
numerous illuminating discussions.

\end{document}